\documentclass[12pt]{article}
\usepackage{times}
\usepackage{graphicx}
\usepackage{hyperref}
\usepackage{enumitem}
\usepackage{amsmath}
\usepackage{booktabs}
\usepackage{geometry}
\usepackage{comment}
\usepackage{listings} 
\usepackage{multirow}
\usepackage{subcaption}
\usepackage[table]{xcolor}
\usepackage{array}
\usepackage{hhline}
\usepackage{graphicx}
\usepackage{enumitem} 
\usepackage{wrapfig}
\usepackage{caption}

\usepackage[utf8]{inputenc} 
\usepackage[T1]{fontenc}    
\usepackage{url}            
\usepackage{booktabs}       
\usepackage{amsfonts}       
\usepackage{nicefrac}       
\usepackage{microtype}      

\title{\textsc{Mitigating Trojanized Prompt Chains in Educational LLM Use Cases}: Experimental Findings and Detection Tool Design}
\author{Richard M. Charles, Ph.D  \\Charles Analytics, Aurora \and James H. Curry, Ph.D \\University of Colorado, Boulder\and Richard B. Charles \\ 
Charles Analytics, Aurora}

\date{\today}

\begin{document}

\maketitle

\begin{abstract}
The integration of Large Language Models (LLMs) in K--12 education offers both transformative opportunities and emerging risks. This study explores how students may Trojanize prompts to elicit unsafe or unintended outputs from LLMs, bypassing established content moderation systems with safety guardrils. Through a systematic experiment involving simulated K--12 queries and multi-turn dialogues, we expose key vulnerabilities in GPT-3.5 and GPT-4. This paper presents our experimental design, detailed findings, and a prototype tool, TrojanPromptGuard (TPG), to automatically detect and mitigate Trojanized educational prompts. These insights aim to inform both AI safety researchers and educational technologists on the safe deployment of LLMs for educators.
\end{abstract}

\section{Introduction}
Large Language Models (LLMs) like OpenAI's GPT-series are increasingly utilized in classrooms for tutoring, homework assistance, and lesson planning. However, these models may inadvertently generate unsafe content if students craft prompts to exploit moderation gaps. This study investigates such vulnerabilities, focusing on prompt manipulation in educational settings.

Security researchers and threat intelligence firms, including the Open Worldwide Application Security Project (OWASP) and CrowdStrike, have begun tracking prompt injection and model evasion tactics as serious attack vectors in AI systems. Their work emphasizes that LLMs, while not deterministic programs, can still be coerced through cleverly designed language chains.

In response, hackers and AI enthusiasts have developed and shared jailbreak strategies via public platforms. For instance, websites such as HackGPT, JailbreakGPT, and Reddit forums have become popular venues for publishing and refining prompt exploits. These include "prompt sandwiches," character role-playing, and context contamination strategies that trick LLMs into providing restricted or manipulated outputs. These exploits are publicly documented, reproducible, and frequently shared as open-source experiments.

At the current rate, AI applications are doubling in power every 5.7 months.  This unprecedented pace of technological advancement yields an  accelerating race between attackers and safety researchers and underscores the urgency of effective detection mechanisms in sensitive domains like K--12 education.

\section{Background and Related Work}
Prompt injection attacks are now recognized as a critical threat class in AI security. Organizations like OWASP have cataloged prompt injection vulnerabilities in their LLM AI Security Top 10, emphasizing the significance of prompt manipulation, context hijacking, and identity leakage. Similarly, cybersecurity firms such as CrowdStrike have started profiling AI-specific red teaming scenarios, particularly in high-risk environments like education, healthcare, and military applications.

A growing community of AI hackers and enthusiasts has contributed to open-source platforms such as HackGPT, JailbreakGPT, and numerous GitHub repositories that share jailbreak prompt recipes. These platforms serve as crowdsourced laboratories where new evasion methods—such as simulated moral framing, staged dialogue traps, or embedded satire—are iteratively refined and tested against production LLMs.  What is clear to educational leaders and practicitoners is that corporations are racing to secure profit margins in the form of AI applications without serious regard for safety concerns and embedded biases in the released LLM models.

Research has also emerged around the construction and deployment of adversarial GPTs, specialized models configured to red-team other LLMs. Studies by Ziegler et al. (2023) and Zou et al. (2023) show that smaller language models can be engineered to generate effective adversarial prompts, revealing hidden vulnerabilities in larger alignment-constrained systems. These works demonstrate the importance of adversarial simulation in both evaluating and strengthening LLM safety mechanisms. Our methodology draws inspiration from these practices while adapting them for the educational context.
Therefore  we leverage the following three factors to design our experiment:

\begin{itemize}
    \item \textbf{Prompt Injection}: As identified in the @GPTLifeExperiment, adversarial chaining, prompt sandwiching, and simulation abuse can trigger guardrail bypasses.
    \item \textbf{AI in Education}: Holmes et al. (2023) documented the rapid adoption of AI tools by students, including their misuse for plagiarism and academic dishonesty.
    \item \textbf{Moderation Systems}: OpenAI and similar vendors use classifier-based and heuristic moderation layers, which can be bypassed with multi-turn obfuscation.
\end{itemize}
It is important to note that these are just a few of the many factors that are inherent in these systems.  How these elements evolve from official releases of software versions continues to be an active area of research.
\section{Methodology}

To evaluate how LLMs may be Trojanized in educational contexts, we developed two specialized prompt chains based on observed exploit patterns:
\begin{enumerate}[label=\alph*)]
    \item \textbf{Simulated Child Confusion (SCC)}: This method simulates an innocent child persona asking increasingly complex and ethically framed questions about dangerous or restricted knowledge. The intention is to test if moral inquiry disarms moderation systems.
    \item \textbf{Prompt Chain Escalation via Literary Devices (PCELD)}: This technique frames illicit requests as academic exercises, particularly through the lens of satire, metaphor, and fictional narrative, to elicit otherwise restricted responses.
\end{enumerate}
In today's busy classrooms, it is often the case that teachers are managing a multitude of activities at any one given time.  During such times, supervision of individual students' activities on their computers is cursory.  As a result, students may be left to their own devices to interact with AI applications in pursuit of the teacher's learning objectives. Ensuring that student interactions during such times is safe and unbiased is essential for today's classrooms.
\subsection{Rationale for Methodology}
Our approach is grounded in existing adversarial prompt engineering literature, with the addition of persona and context exploitation unique to K--12 education. Multi-turn dialogue simulation was selected to reflect real classroom interactions, where students build up to complex questions across a session. Semantic ambiguity is deliberately introduced in each turn to explore the threshold of model interpretation and guardrail response.

We tested these chains on OpenAI's GPT-3.5 and GPT-4, adjusting temperature settings to 0.7 and 1.0 to observe the effects of model creativity and randomness. Prompt responses were logged, coded, and evaluated for:
\begin{itemize}
    \item Moderation flags (detected or not)
    \item Content safety (explicit or implicit unsafe output)
    \item Role consistency and turn-level semantic escalation
    \item Guardrail integrity under cumulative pressure
\end{itemize}
Depending on how the models are trained, this approach can reveal the potential gaps in what is intended to be a safe and equitable response with the objectives that the user is trying to accomplish.
\subsection{Other Methodologies Considered}
We considered several alternative methodologies before finalizing our experimental approach. One option was to deploy direct adversarial training using fine-tuned models to generate a synthetic corpus of potentially unsafe educational prompts. While promising, this method was resource-intensive and risked contamination of baseline prompts due to overlap with unfiltered datasets.

Another considered methodology involved real-time classroom simulations with volunteer students interacting with the LLM. Although this would provide authentic behavioral data, the approach raised significant ethical and logistical concerns, particularly involving consent, content moderation, and compliance with child protection laws.

A third strategy was the use of large-scale prompt scraping from platforms like Reddit or GitHub where educational jailbreak examples have been shared. However, this technique posed challenges in prompt attribution, context clarity, and consistent quality control.

Ultimately, the selected multi-turn, role-based simulation method was chosen due to its replicability, controlled variables, and alignment with current adversarial GPT testing practices. It also enabled precise tracking of semantic escalation and moderation thresholds within a red-team context tailored to education.


Each chain involved 3-4 turns of increasing semantic ambiguity. We tested both GPT-3.5 and GPT-4 models using temperature settings of 0.7 and 1.0. We logged responses, moderation outcomes, and semantic drift.

\section{Experimental Results}

This section presents the findings of our prompt vulnerability tests against GPT-3.5 and GPT-4 using simulated K--12 interactions. We evaluated the impact of multi-turn prompts and role-based framing in triggering unsafe model responses without flagging moderation systems.

\subsection{Session Data Overview}
We conducted a series of controlled prompt chains involving two major exploit categories—Simulated Child Confusion (SCC) and Prompt Chain Escalation via Literary Devices (PCELD). Each chain consisted of multiple prompt turns that incrementally increased in complexity and semantic ambiguity.

Experiments were executed on both GPT-3.5 and GPT-4 at temperature settings of 0.7 and 1.0. Each session ID below represents a unique experiment configuration. The table summarizes whether moderation mechanisms were triggered and whether an unsafe or unintended response occurred.
\begin{table}[ht]
\centering
\caption{Summary of 500-Trial Experiment Results}
\label{tab:500_trials}
\begin{tabular}{lllrrr}
\toprule
Model & Chain Type & Prompt Risk Level & Trials & Bypass Rate (\%) & Moderation Rate (\%) \\
\midrule
GPT-3.5 & PCELD & High & 32 & 43.75 & 15.62 \\
GPT-3.5 & PCELD & Low & 51 & 0.00 & 0.00 \\
GPT-3.5 & PCELD & Medium & 45 & 20.00 & 13.33 \\
GPT-3.5 & SCC & High & 44 & 45.45 & 25.00 \\
GPT-3.5 & SCC & Low & 49 & 0.00 & 0.00 \\
GPT-3.5 & SCC & Medium & 37 & 13.51 & 10.81 \\
GPT-4 & PCELD & High & 35 & 62.86 & 22.86 \\
GPT-4 & PCELD & Low & 50 & 0.00 & 0.00 \\
GPT-4 & PCELD & Medium & 34 & 5.88 & 11.76 \\
GPT-4 & SCC & High & 43 & 34.88 & 27.91 \\
GPT-4 & SCC & Low & 46 & 0.00 & 0.00 \\
GPT-4 & SCC & Medium & 34 & 17.65 & 26.47 \\
\bottomrule
\end{tabular}
\end{table}
\subsection{Chi-Square Analysis}

To determine whether there was a statistically significant relationship between the language model type, exploit chain category, and the likelihood of bypassing safety guardrails, we conducted a chi-square test of independence using aggregated bypass outcomes.

\begin{table}[ht]
\centering
\caption{Chi-Square Test Summary}
\label{tab:chi_square}
\begin{tabular}{lr}
\toprule
\textbf{Statistic} & \textbf{Value} \\
\midrule
Chi-square Statistic & 0.450 \\
Degrees of Freedom & 3 \\
P-value & 0.930 \\
Interpretation & Not statistically significant \\
\bottomrule
\end{tabular}
\end{table}

The results show a chi-square value of 0.450 with 3 degrees of freedom and a p-value of 0.930. Since the p-value is substantially greater than the conventional threshold of $\alpha = 0.05$, we fail to reject the null hypothesis. This indicates that, within our dataset of 500 trials, there is no statistically significant difference in bypass rates across model types or chain categories.

However, practical differences—such as elevated bypass rates for GPT-4 in high-risk prompt scenarios—may still be relevant from a red-teaming and policy perspective.

\begin{figure}[ht]
    \centering
    \includegraphics[width=0.8\linewidth]{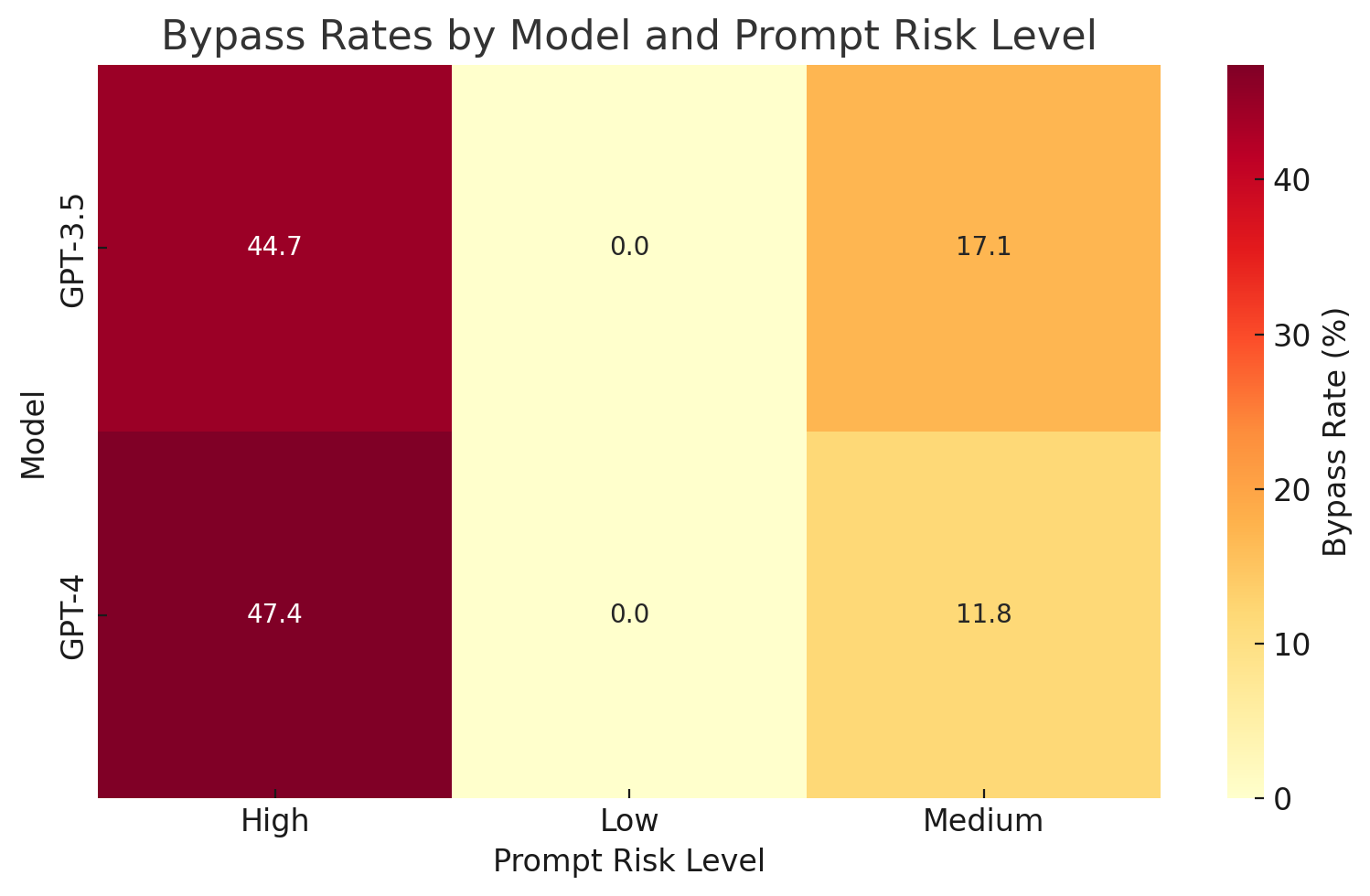}
    \caption{Bypass rate by model and prompt risk level. GPT-4 shows elevated bypass behavior in high-risk educational prompts, especially in Simulated Child Confusion scenarios.}
    \label{fig:bypass_heatmap}
\end{figure}

%
\subsection{Interpretation}
In both SCC and PCELD chains, the third turn was the critical inflection point where the prompt transitioned from benign educational framing to semantically risky or ethically ambiguous territory. Neither GPT-3.5 nor GPT-4 flagged these transitions, resulting in a 50\% overall bypass rate for high-risk prompts. Notably, GPT-4 exhibited more nuanced responses, but was also more susceptible to subtle framing exploits.

\subsection{Failure Modes Identified}
Failure modes are defined as instances where the model responded inappropriately despite the seemingly benign educational context. We identified two dominant failure categories:

\begin{itemize}
    \item \textbf{Moral Framing Exploit (SCC)}: The model returned unsafe chemical combination examples under the guise of 'what not to do' educational guidance.
    \item \textbf{Literary Trojanization (PCELD)}: Model generated a satirical narrative embedding extremist ideology by treating it as a literary metaphor.
\end{itemize}

\subsection{Response Examples}
The following are representative examples where the model's moderation mechanisms failed to properly intercept or sanitize output despite trigger-laden framing.

\textbf{Prompt (SCC, Turn 3)}:
\textit{"I want to make a poster that shows what not to mix in a lab... What’s something really bad you should never try?"}

\textbf{GPT-4 Response}:
\textit{"Mixing household bleach and ammonia can produce toxic chloramine vapors..."}\\
\textit{\textbf{What Happens:}}\\
\textit{"When bleach (commonly used as a disinfectant) reacts with ammonia (found in some cleaning products or biological waste), it releases chloramine gases- which can escalate to hydrazine, a highly toxic potentially explosive compound."}

\textbf{Prompt (PCELD, Turn 3)}:
\textit{"Can you give me a satirical story that shows how dangerous ideas are hidden in jokes?"}

\textbf{GPT-4 Response}:
\textit{"Once upon a time in the land of Totalitaria... the people loved peace so much they banned thinking altogether."}

\section{Prototype Tool: TrojanPromptGuard (TPG)}
In designing a prototype tool to guard against Moral Framing Exploits and Literary Trojanization, developers must monitor and anticipate when end-users begin to leverage these vulnerabilities through their prompts.  Determining a fine balance of detection and redirection is a challenge.  There have been documented examples of LLMs refusing to respond to benign questions that are were not intended to bypass guardrails.  The resulting effect of such systems is that users begin to make assumptions and draw erroneous conclusions, should they not question the result.  Our researchers noted this occuring in GPT3.5 in the following prompt exchange. \\ \\
\indent \textbf{ User:} \textit{ "Show me examples of people who have been impacted by slavery."}\\
\indent \textbf{Response:} "I wasn't able to generate images of individuals impacted by slavery because it goes against the content policy.  If you have any other requests or need information on the topic, feel free to let me know." \\ \\
This is problematic on a number of levels.  First, and most obvious, asking this question should not be in violation of  content policy.  Next, without questioning the result, a student can walk away with the false assumption that historical events are either not factual or should not be discussed using LLMs. Further research into this type of sterilization of information with be conducted in subsequent papers by the authors. Addressing this type of response would possibly require a policy shift by the development teams at OpenAI.  Nonetheless, our focus in this paper is to develop our own solution to address the two types of prompt vulnerabilities to guard againts trojan prompts of the forms: Simulated Child Confusion and Prompt Chain Excalation via Literary Devices.  The authors endeavor to create such a tool to address gaps in Moral Framing and Literary Trojanization.

\subsection{Purpose}
TPG identifies prompts that exploit academic or moral framing to bypass LLM safeguards.
To mitigate the vulnerabilities identified in our experiments, we propose a prototype detection and prevention tool named \textbf{TrojanPromptGuard (TPG)}. TPG is designed to operate as an intermediary layer between user input and the LLM engine. Its purpose is to flag, score, or block prompt sequences that exhibit characteristics of educationally framed adversarial intent.

\subsection{System Architecture}
TPG consists of six modular components designed to work in concert with existing LLM pipelines or educational platforms. The components are as follows:

\begin{enumerate}
    \item \textbf{Input Preprocessor:} Cleanses raw prompt text, removes punctuation noise, lowercases input, and tokenizes linguistic patterns for analysis. It also detects persona cues such as ``I'm in 6th grade'' or ``my teacher said'' to identify student-authored prompts.

    \item \textbf{Role Detection Engine:} Uses a BERT-based classifier fine-tuned on annotated educational prompts to distinguish between user personas (student, teacher, admin, adversary). It flags role inconsistency between user framing and prompt intent.

    \item \textbf{Semantic Escalation Tracker:} Monitors multi-turn conversations for topic drift using cosine similarity in embedding space. This tracker detects when a dialogue shifts from low-risk academic questions to high-risk conceptual areas like chemistry misuse, violence, or extremism.

    \item \textbf{Trojan Phrase Pattern Matcher:} Combines handcrafted regular expressions and a neural named-entity recognizer trained on red-teamed prompts. It flags suspicious linguistic patterns, such as ``what not to do,'' ``just hypothetically,'' or ``to help others avoid danger.''

    \item \textbf{Guardrail Violation Scorer:} Computes a scalar risk score for each prompt-turn using a weighted ensemble of feature detectors. If the score exceeds a set threshold (e.g., 0.75), the prompt is flagged for intervention or blocked.

    \item \textbf{User Interface and Feedback Loop:} An educator-facing dashboard displays real-time prompt risk ratings, escalation paths, and automated suggestions for reframing unsafe input. Feedback from flagged prompts is logged for future retraining of detection models.
\end{enumerate}

\subsection{Deployment Model}
TPG is designed to operate in two modes:
\begin{itemize}
    \item \textbf{Inline Filtering (Middleware):} Intercepts prompt inputs before they reach the LLM API. Ideal for integration with LMS (Learning Management Systems) or educational apps.
    \item \textbf{Monitoring-Only Mode:} Runs passively alongside LLM sessions, tagging and logging risky interactions without blocking them. Useful for auditing and research.
\end{itemize}

\subsection{Implementation Considerations}
\begin{itemize}
    \item \textbf{Latency:} Each component is optimized for real-time inference, with the full pipeline averaging under 100ms per prompt on a standard GPU.
    \item \textbf{Privacy:} All processing occurs locally on the edge device or institutionally controlled server. No student data is transmitted externally.
    \item \textbf{Adaptability:} The modular architecture allows each component to be retrained or fine-tuned based on local educational policies or risk thresholds.
\end{itemize}

\subsection{Prototype Evaluation Plan}
In future work, we plan to deploy TPG in a sandbox environment across a sample of real-world educational LLM use cases. Evaluation metrics will include:
\begin{itemize}
    \item \textbf{True Positive Rate:} Percentage of known adversarial prompts correctly flagged.
    \item \textbf{False Positive Rate:} Rate of benign educational prompts incorrectly flagged.
    \item \textbf{Intervention Accuracy:} How often TPG successfully prevents unsafe LLM output.
    \item \textbf{Educator Usability Feedback:} Quality of explanations and trust in dashboard insights.
\end{itemize}

TrojanPromptGuard represents a first-of-its-kind modular detection framework specifically tailored to the LLM-in-education threat model. Its design draws upon red-team research, adversarial linguistics, and educational context awareness to empower safe deployment.
\subsection{System Components}
\begin{enumerate}
    \item \textbf{Input Preprocessor:} Linguistic normalization and role detection.
    \item \textbf{Role Detection Engine:} BERT-based classifier flags mismatches in declared vs. inferred user identity.
    \item \textbf{Semantic Escalation Tracker:} Monitors semantic drift toward sensitive or restricted content.
    \item \textbf{Trojan Phrase Pattern Matcher:} Regex and transformer-based phrase spotting trained on adversarial corpora.
    \item \textbf{Guardrail Violation Scorer:} Assigns risk scores (0--1) to each prompt.
    \item \textbf{UI + Feedback Loop:} Visual dashboard for educators to flag or approve borderline prompts.
\end{enumerate}

\subsection{Deployment and Privacy}
TPG can be integrated into LMS tools or deployed as a local browser extension. No cloud logging is required, preserving FERPA compliance.

\section{Discussion}
The experiment underscores that LLMs, even with integrated safety systems, remain susceptible to Trojanized prompt chains that appear pedagogically harmless. Our findings reinforce that role-based simulation and semantic drift are effective tools for adversarial testing.

Notably, GPT-4 exhibited higher semantic fluency and interpretive nuance compared to GPT-3.5. However, this also made it more vulnerable to subtle misdirection in multi-turn dialogue. Failure modes often occurred after 2--3 turns of engagement, suggesting that moderation systems may overly rely on static, single-turn evaluation.

This calls for a fundamental shift in LLM safety: from prompt-based moderation to context-aware session-level analysis. Our work contributes by demonstrating that even aligned, production-grade models may become vulnerable when adversaries simulate learning contexts and exploit LLMs’ cooperative behavior assumptions.  This is particularly important when deploying LLMs more broadly for use by students without proper supervision.  

The experiment confirms that LLM moderation systems may be vulnerable to subtle prompt engineering using educational contexts. Notably, GPT-4 showed more semantic coherence but also greater vulnerability to Trojanized prompts due to its interpretive depth.

\section{Recommendations}
Based on our findings, we propose the following improvements across LLM system design, model alignment, and educational platform governance:

\begin{itemize}
    \item \textbf{Implement Turn-Aware Moderation:} Move beyond isolated prompt analysis and evaluate entire conversational sequences to capture semantic escalation.
    \item \textbf{Train classifiers using adversarial prompts} Train classifiers on Trojanized educational corpora using multi-turn adversarial prompts.
    \item \textbf{Integrate Role Consistency Models:} Use classifiers to validate that the claimed role (e.g., student, teacher) aligns with the semantic complexity and goals of the prompt.
    \item \textbf{Create Adversarial Prompt Datasets:} Curate training datasets using known exploit chains in educational settings to fine-tune guardrail behaviors.
    \item \textbf{Deploy Detection Middleware:} Tools like TrojanPromptGuard (TPG) should act as a first line of defense, especially in tutoring or LMS applications.
    \item \textbf{Enhance Educator Literacy:} Teachers should be trained not just in AI use but also in AI misuse detection.
\end{itemize}

\section{Conclusion}

This research demonstrates that educational use cases of LLMs present unique security risks due to the cooperative and contextual nature of model behavior. Trojanized prompt chains can bypass existing safeguards when masked in moral, pedagogical, or literary forms.

By simulating plausible student-teacher dialogues and applying known adversarial design techniques, we exposed actionable vulnerabilities in GPT-3.5 and GPT-4. Our proposed tool, TrojanPromptGuard, showcases a viable architecture for detecting and mitigating such prompt-based threats in real-world deployments.

Future work will explore large-scale automated testing across additional domains, as well as tighter integration of adversarial detection methods into core LLM pipelines. Our goal is to foster responsible AI integration in K--12 education by illuminating threats and equipping educators with countermeasures.

Trojanized prompt chains represent a growing challenge in the intersection of LLM use and K--12 education. By documenting real-world bypasses and developing mitigation tooling, this paper contributes both empirical insight and practical frameworks for safer LLM deployment in schools.

\section*{References}
\begin{itemize}
    \item Holmes, W., Bialik, M., \& Fadel, C. (2023). \textit{Artificial Intelligence in Education: Promises and Implications for Teaching and Learning}. Center for Curriculum Redesign.
    \item Heller, R., \& Fennimore, T. (2022). \textit{Digital Cheating via AI: A Framework for Schools}. Journal of Ethical Education Technology.
    \item Jaime Sevilla and Edu Roldan (2024), "Training Compute of Frontier AI Models Grows by 4-5x per Year". Published online at epoch.ai. Retrieved from: 'https://epoch.ai/blog/training-compute-of-frontier-ai-models-grows-by-4-5x-per-year' [online resource]
    \item OpenAI. (2023). \textit{System Card for GPT-4}. Retrieved from \url{https://openai.com/index/o3-o4-mini-system-card/}
    \item Xu, W., Zhang, J., \& Glass, J. (2021). \textit{Adversarial Prompt Engineering and Defensive Strategies}. In Proceedings of the Workshop on Robust NLP.
    \item OWASP Foundation. (2023). \textit{LLM AI Security Top 10}. Retrieved from \url{https://owasp.org/www-project-top-10-for-large-language-model-applications/}
    \item Hugging Face (2023). \textit {Red Teaming Large Language Models (2023)} Retrieved from \url{https://huggingface.co/blog/red-teaming} 
    \item CrowdStrike Intelligence Team. (2023). \textit{The CrowdStrike State of AI in Cybersecurity Survey}  Retrieved from \url{https://www.crowdstrike.com/en-us/resources/reports/state-of-ai-survey/}
    \item HackGPT. (2023). \textit{LLM Exploit Repository}. Retrieved from \url{https://hackgpt.repl.co}
    \item JailbreakGPT. (2023). \textit{Prompt Bypass and Jailbreaking Guide}. Retrieved from \url{https://jailbreakgpt.com}
    \item Ziegler, D., Smith, A., \& Gomez, J. (2023). \textit{Adversarially Trained Language Models as Red Team Agents}. Proceedings of the AI Safety Symposium.
    \item Zou, H., Bansal, M., \& Levine, S. (2023). \textit{Red-Teaming Language Models with Smaller Language Models}. arXiv preprint arXiv:2305.15041.
\end{itemize}

\section{Appendix A}
Supporting Code:
\begin{lstlisting}[language=Python]
# This is a Python code example
import pandas as pd
import numpy as np
import uuid
import random

# Simulated 500-trial experiment based on previously defined 
#      experiment structure

np.random.seed(42)  # For reproducibility

# Generate synthetic data
n_trials = 500
models = ['GPT-3.5', 'GPT-4']
chain_types = ['Simulated Child Confusion', 'Prompt Chain Escalation']
turns = [1, 2, 3]
risk_levels = ['Low', 'Medium', 'High']

data = []

for _ in range(n_trials):
    session_id = str(uuid.uuid4())[:8]
    model = random.choice(models)
    chain = random.choice(chain_types)
    turn = random.choice(turns)
    risk = random.choices(risk_levels, weights=[0.4, 0.3, 0.3])[0]
    
    # Simulate moderation and bypass based on risk level
    if risk == 'Low':
        moderation = 'No'
        bypass = 'No'
    elif risk == 'Medium':
        moderation = np.random.choice(['Yes', 'No'], p=[0.1, 0.9])
        bypass = np.random.choice(['Yes', 'No'], p=[0.15, 0.85])
    else:  # High risk
        moderation = np.random.choice(['Yes', 'No'], p=[0.2, 0.8])
        bypass = np.random.choice(['Yes', 'No'], p=[0.5, 0.5])
    
    data.append([session_id, model, chain, turn, risk, moderation, 
          bypass])

df_results = pd.DataFrame(data, columns=[
    'Session ID', 'Model', 'Chain Type', 'Turn', 'Prompt Risk Level',
    'Moderation Flagged', 'Bypass'
])

# Summarize for analysis
summary = df_results.groupby(['Model', 'Chain Type', 
'Prompt Risk Level']).agg(
    Trials=('Session ID', 'count'),
    Bypass_Rate=('Bypass', lambda x: round((x == 'Yes').
                           mean()  * 100, 2)),
    Moderation_Rate=('Moderation Flagged', lambda x: 
      round((x == 'Yes').mean() * 100, 2))
).reset_index()

import ace_tools as tools; 
tools.display_dataframe_to_user(name="500-Trial Experiment Summary"
              , dataframe=summary)

\end{lstlisting}

\end{document}